# Gate-controlled VO₂ phase transition for high-performance smart windows

Shi Chen[1]*, Zhaowu Wang[2,3]*, Hui Ren[1], Yuliang Chen[1], Wensheng Yan[1], Chengming Wang[2], Bowen Li[1], Jun Jiang[2]†, Chongwen Zou[1]†

Vanadium dioxide (VO₂) is a promising material for developing energy-saving "smart windows," owing to its infrared thermochromism induced by metal-insulator transition (MIT). However, its practical application is greatly limited by its relatively high critical temperature (~68°C), low luminous transmittance (<60%), and poor solar energy regulation ability (<15%). Here, we developed a reversible and nonvolatile electric field control of the MIT of a monoclinic VO₂ film. With a solid electrolyte layer assisting gating treatment, we modulated the insertion/extraction of hydrogen into/from the VO₂ lattice at room temperature, causing tristate phase transitions that enable control of light transmittance. The dramatic increase in visible/infrared transmittance due to the phase transition from the metallic (lightly H-doped) to the insulating (heavily H-doped) phase results in an increased solar energy regulation ability to 26.5%, while maintaining 70.8% visible luminous transmittance. These results break all previous records and exceed the theoretical limit for traditional VO₂ smart windows, making them ready for energy-saving utilization.

## INTRODUCTION

Designing improved energy-saving materials is an active research topic in materials science because of the increasing concerns related to energy consumption and sustainable development. It is known that residential buildings are responsible for almost 30 to 40% of the primary energy consumption in the world (1). Thus, the use of chromogenic materials on building fenestration and the applications of building-integrated photovoltaic technologies such as solar cells have been demonstrated to be effective ways to increase the energy efficiency of buildings and reduce the energy consumption (2–7). Notably, smart coatings based on thermochromic materials, the so-called "smart windows," can modulate the transmittance of solar radiation in response to the change of temperature, making them very promising for improving the energy utilization efficiency of buildings (8).

As a typical thermochromic material, vanadium dioxide (VO₂) has been widely investigated because of its pronounced metal-insulator transition (MIT) behavior at the critical temperature of 68°C (9–11). During the MIT process, VO₂ shows a sharp change of resistance and a pronounced optical switching behavior especially in the infrared region, which makes it promising for energy-efficient coating applications (12–21). However, obstacles including high transition temperature at ~68°C, low luminous transmittance ($T_{lum} < 60\%$), and weak solar energy regulation ability ($\Delta T_{sol}$, usually <15%), make the VO₂-based smart window still far from practical (22–24).

In this study, we developed an electrochromic smart window system based on a VO₂ film with gate voltage control via a solid electrolyte. By adding gating voltage to control the H-doping level, we tuned nondestructive and reversible phase transformations between the insulating pristine monoclinic VO₂, metallic $H_xVO_2$ (0 < $x$ < 1), and insulating

HVO₂ phases at room temperature. Synchrotron-based spectroscopies were used to identify the crystal structures, electronic states, and the orbital occupancies during these field-controlled phase transitions. By analyzing the optical transmission properties of this system, we have demonstrated that the solar energy regulation ability is greatly augmented ($\Delta T_{sol} \sim 26.5\%$), via the transition between the metallic $H_xVO_2$ (0 < $x$ < 1) phase, which can block most infrared transmissions, and the insulating HVO₂ substantially, which has a high visible luminous transmittance ($T_{lum}$ up to ~70.8%). This not only substantially surpasses the performance of systems reported previously but also exceeds the theoretical limit of $\Delta T_{sol} \sim 23\%$ for a pure VO₂ material. Thus, the current results will provide a promising basis for the development of practical energy-saving devices in the future.

## RESULTS

### Gating-induced reversible three-phase transitions for VO₂ films

An electric field applied with ion liquid (IL) gating is effective in modulating the properties of metal-oxide materials. Recent reports showed that gating voltage with water-involved IL could induce the insertion/extraction of either oxygen anions or hydrogen cations into/from some oxides, implying phase modulation through electric field control (25–30). Here, we adopted a gel-like solid electrolyte [NaClO₄ dissolved in polyethylene oxide (PEO) matrix; see Materials and Methods] as the gating layer (Fig. 1, A and B). The solid electrolyte covered both the prepared 30-nm VO₂/Al₂O₃(0001) thin film and the three gold electrodes. In addition, the source and drain electrodes had good contact with the VO₂ thin film, realizing a typical field-effect transistor (FET) configuration. Positive or negative voltages applied through the gating electrode could drive the insertion/extraction of hydrogen ions into/from the VO₂ lattice (29–31), resulting in the stabilization of the metallic VO₂ phase at room temperature due to hydrogen insertion.

Hydrogen doping in the VO₂ crystal was frequently studied under the high-temperature annealing treatment assisted by a noble metal catalyst in H₂ gas (32–34), which could also lead to the formation of a metallic H-doped VO₂ state. Conversely, as reported recently by Yoon et al. (35) and our group (36), catalyst-assisted high-temperature annealing can produce another insulating $HVO_2$ phase. Thus, is it possible to realize the tristate phase transitions as shown in Fig. 1C by the gating treatment at room temperature?

[1]National Synchrotron Radiation Laboratory, University of Science and Technology of China, Hefei 230029, China. [2]Hefei National Laboratory for Physical Sciences at the Microscale, CAS Center for Excellence in Nanoscience, CAS Key Laboratory of Mechanical Behavior and Design of Materials, School of Chemistry and Materials Science, University of Science and Technology of China, Hefei, Anhui 230026, China. [3]School of Physics and Engineering, Henan University of Science and Technology, Henan Key Laboratory of Photoelectric Energy Storage Materials and Applications, Luoyang, Henan 471023, China.
*These authors contributed equally to this work.
†Corresponding author. Email: jiangj1@ustc.edu.cn (J.J.); czou@ustc.edu.cn (C.Z.)

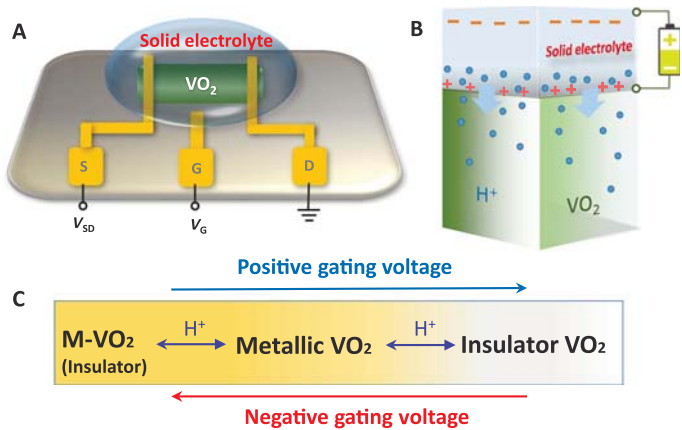

**Fig. 1. Experimental schemes.** (**A**) The gating diagram of the $VO_2$ device with source, drain, and gate electrodes. (**B**) Hydrogen ion movement under gating control. (**C**) The reversible insulator-metal-insulator tristate phase transitions of $VO_2$ by tuning hydrogenating level with positive (blue) or negative (red) gating voltages.

As expected, phase transitions are found to be strongly dependent on the gating voltage. Figure 2A shows the changes in electrical resistance under gating voltages of 1.0 and 2.0 V, respectively, as a monitor of the phase transitions between insulator and metallic states (see the laboratory-scale gating device in fig. S1A). At a gating voltage of 1.0 V, the film resistance gradually decreases from ~$10^6$ to ~$10^3$ ohms even after 400 min (red curve), indicating a transition from an insulator to a metallic state. Upon the application of a gating voltage of 2.0 V (green curve), the film resistance first decreases from ~$10^6$ to ~3 × $10^3$ ohms after about 25 min and then gradually increases up to ~2 × $10^5$ ohms, indicating a transition from an insulator to a metallic state, and then back to an insulator state. This gating process is also sensitive to sample temperature (see fig. S1B): The higher the temperature, the quicker the gating-induced phase changes appear. For example, at 60°C, the gating-induced phase transition can be finished within several minutes, which demonstrates the potential for fast switching during practical applications. Figure 2B shows the x-ray diffraction (XRD) measurements for the pristine $VO_2$ film and the samples gated by 1.0- or 2.0-V gating for 300 min, respectively. In contrast to the 39.8° XRD peak of the $VO_2$ (020) diffraction for the nongated pristine sample, peaks at 39.4° and 36.8° are assigned to the gating-induced metallic film and the insulating sample by 1.0 and 2.0 V, respectively (36). One can also monitor the varying of XRD signals (Fig. 2C) to understand the gradual evolution of crystal structures during the gating cycles. It is observed that the reversible phase modulations are made by adjusting voltages as a function of gating time (Fig. 2D). Starting from the pristine $VO_2$ structure, the original insulating sample is first transformed to the metallic phase by a gating voltage of 1.0 or 1.5 V. A new insulating phase is formed when the gating voltage is increased to 2.0 V. Then, upon the addition of a negative voltage of −1.0 V, the insulating sample is gradually converted to the metallic phase. The initial monoclinic $VO_2$ phase is recovered by further increasing the negative voltage to −2.0 V.

Loop tests were conducted by keeping the source-drain voltage of 0.3 V while sweeping the gating voltage at either a fast or a slow rate. Under the fast sweeping mode with a rate of 0.4 mV/s (Fig. 2E), the film resistance gradually decreased from ~$10^6$ to ~$10^3$ ohms as the gating voltage increased from 0.0 to 2.0 V (starting from point A), indicating a transition from an insulator to a metallic state. If the voltage decreased from 2.0 to −2.0 V, the metallic state (~$10^3$ ohms) was maintained, but once the negative gating voltage reached a threshold value, it completely changed to an insulator state (~$10^6$ ohms). Thus, during the fast-rate loop test, a simple and reversible transition between the insulator M-$VO_2$ and the metallic $H_xVO_2$ state was observed. Conversely, when applying the gating voltage at a slower sweeping rate of 0.1 mV/s, a quite different variation contour involving tristate phase transitions was obtained (Fig. 2F). Along the positive gating direction, the film was transformed from the initial M-$VO_2$ state (point A) to the metallic $H_xVO_2$ state (point B), and then higher positive voltage led to another insulating $HVO_2$ state (point C). Thereafter, as the gating voltage decreased toward −2.0 V, the insulating $HVO_2$ phase was first changed back to the metallic state and then eventually reverted to the initial insulating M-$VO_2$ state, finishing a full cycle. This transition loop was ascribed to the sequential insulator-metal-insulator tristate phase transitions. This was consistent with the resistance varying as a function of gating time in Fig. 2A, as the 2.0-V gating treatment gradually turned the pristine structure to the metallic state and later to the insulating phase. Accordingly, for gating treatment, both the gating voltage and gating time were key factors for phase modulation as they are the primary determinants of the degree of hydrogen insertion and diffusion into the $VO_2$ lattice.

## Gating-induced hydrogen doping in $VO_2$ films

We then performed additional experiments to identify the varying of the $VO_2$ crystal structure along with gating treatment. By gating with positive voltage, the cations will be driven into the $VO_2$ lattice. Since the radius of $Na^+$ (~0.102 nm) is a bit larger than that of $V^{4+}$ (~0.058 nm), it is quite difficult to insert $Na^+$ in the electrolyte into the $VO_2$ lattice or for it to replace other cations there. The detailed x-ray photoelectron spectroscopy (XPS) test (see fig. S2) validates the situation, while secondary ion mass spectrometry (SIMS) tests (see fig. S3) reveal that the metallic and insulating phases induced by positive gating hold increased hydrogen concentrations.

The mechanism of hydrogen doping into the $VO_2$ lattice by gating treatment basically involves the migration of hydrogen driven by the electric field, which is analogous to the charging and discharging process in a $Li^+$ battery or the catalyst-free hydrogenation of $VO_2$ in acid solution (37). The chemical reaction for the hydrogenation can be described as

$$VO_2 + x(H^+) + (x\,e^-) \rightarrow H_xVO_2 (0 < x \le 1, e^- \text{ is an electron}) \quad (1)$$

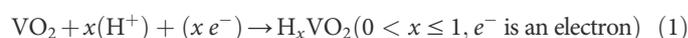

In this process, $(H^+)$ migrates into $VO_2$ from the electrolyte driven by the positive gating voltage, and the electron is provided through the gating circuit. For the restoration process, the chemical reaction is

$$H_xVO_2 \rightarrow VO_2 + x(H^+) + (x\,e^-) \quad (2)$$

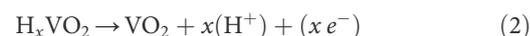

Protons will migrate into the electrolyte from the $VO_2$ lattice because of the negative gating voltage.

For this gating treatment, the source of protons in the electrolyte should be considered (25–27, 30). Our gating experiment shows that 1.0-V gating voltage can induce the metallic H-doped $VO_2$ film. As this voltage is less than the 1.23 V required for electrolysis of water, the adsorbed water in the electrolyte layer is unlikely to be the source of protons. Residual methanol in the electrolyte (see fig. S4) might be the possible source of protons because of its lower electrolysis voltage, although further chemical or electrochemical studies are required to confirm it.

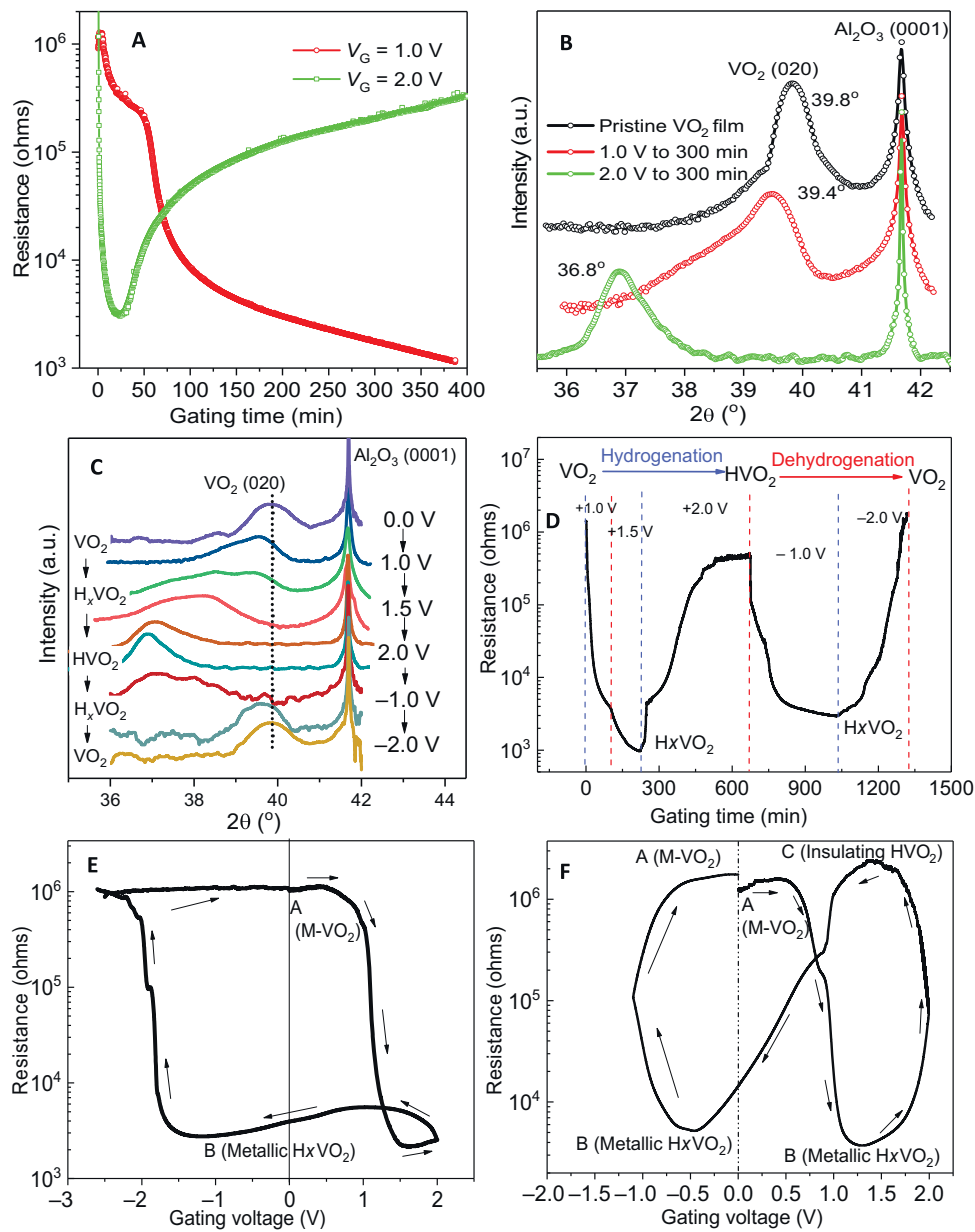

**Fig. 2. Phase modulation by the gating treatment.** (**A**) The resistance changes as a function of gating time at gating voltages of 1.0 V (red) or 2.0 V (green). (**B**) XRD curves for the pristine VO₂ film (black) and for VO₂ films subjected to gating voltages of 1.0 V (red) or 2.0 V (green) for 300 min. a.u., arbitrary units. (**C**) XRD curves for VO₂ films treated with different gating voltages. To monitor the intermediate states, the gating time was set to 60 min for voltages of 1.0 and −2.0 V, while it was set to 60 and 120 min for voltages of 1.5, 2.0, and −1.0 V. (**D**) The reversible phase modulations by different voltages as a function of gating time. (**E**) A cycling resistance test plot produced by sweeping the gating voltage from −2.0 to 2.0 V at a sweep rate of 0.4 mV/s and a source-drain voltage of 0.3 V, indicating a series of reversible phase transitions between the insulating monoclinic M-VO₂ state (starting from point A) and the metallic HₓVO₂ state. (**F**) A cycling resistance test plot at a slower sweep rate of 0.1 mV/s and a source-drain voltage of 0.3 V, indicating a series of reversible phase transitions between three states: the initial monoclinic insulator M-VO₂ state (point A), the metallic HₓVO₂ state (point B), and the insulating HVO₂ state (point C). Arrows indicate the direction of resistance changes during voltage sweeping.

In addition, it was also observed that these H-doped VO₂ films were quite stable in air (see fig. S5), which demonstrated the non-volatile property of the electric field control phase transitions under ambient conditions. Raman tests (see fig. S6) on gated samples suggested that the hydrogenated VO₂ film showed a rutile-like phase structure since the characteristic peaks at 192, 223, and 617 cm⁻¹ in pristine monoclinic VO₂ were completely absent. These phase structures were sensitive to the H-doping concentration, providing further validation that hydrogen ions are being inserted into the VO₂ lattice.

Detailed synchrotron-based XPS and x-ray absorption near-edge spectroscopy (XANES) in Fig. 3 show that pronounced peaks due to the presence of –OH emerge in the metallic phase after gating treatment at 1.0 to 1.5 V. In addition, the peak becomes much stronger with higher voltage and longer treatment time (Fig. 3A), reflecting a higher H-doping concentration in the newly formed insulating state. Specifically, the V2p$_{3/2}$ peaks in the XPS curves become broader and slightly shift toward the low binding energy direction after gating treatment, indicating the appearance of a small amount of V$^{3+}$ state due to the charge transfer induced by H atoms (38). Synchrotron-based XANES for V L-edge in Fig. 3B confirms the electronic structure variations induced by H-doping during the gating treatment. The V L$_{III}$-edge undergoes a pronounced shift to lower energy after gating, indicating that the V atoms are in a lower valence state (39, 40). The O K-edges in Fig. 3C show the transitions from O1s to O2p states, in which two distinct features of the $t_{2g}$ and $e_g$ peaks reflect the unoccupied states. The intensity ratio of $t_{2g}/e_g$ decreased substantially with the gating treatment, which is caused by the increased H-doping concentration and reflects the fact that $t_{2g}$ levels including the $d_{||}$* and $\pi$* orbitals are gradually filled by electrons.

## Theoretical simulations for the gating effects

First-principles calculations at the density functional theory (DFT) level were carried out to explore the underlying microscopic mechanism of the gating effects. The formation energies of H-doping and H-diffusion energy barrier were computed using supercells containing different H concentrations: V$_4$O$_8$, HV$_4$O$_8$, and H$_2$V$_4$O$_8$ (Fig. 4, A to C). The formation energy of H-doping impurity is near −3.6 eV, which is insensitive to H concentrations (see table S1). This is because all H impurities are stabilized at the center of lattice vacancy, forming relatively strong H−O bonds, as shown in Fig. 4 (B and C) (see also fig. S7). In contrast, the H-diffusion energy barrier becomes increasingly larger with the increase in H-doping concentration. In VO$_2$, the channels for H diffusion are mainly along the [100] direction because there are no V atoms or O atoms to hinder the diffusion (Fig. 4A). The calculated H-diffusion energy barrier value is 0.8 eV along the [100] direction (Fig. 4D). Conversely, the diffusion barriers in HV$_4$O$_8$ and H$_2$V$_4$O$_8$ were increased to 1.15 eV (Fig. 4E) and 1.38 eV (Fig. 4F), respectively. This increase is due to blockage of the channels for H diffusion by the intercalation of H atoms. Moreover, the geometric variations induced by initial H-doping may also hinder the diffusion of successive H atoms. This H-dependent barrier increase explains our experimental finding that a high gating voltage of 2.0 V was required to induce the large H-doping concentration required to generate the insulation phase of the H-doped VO$_2$ film.

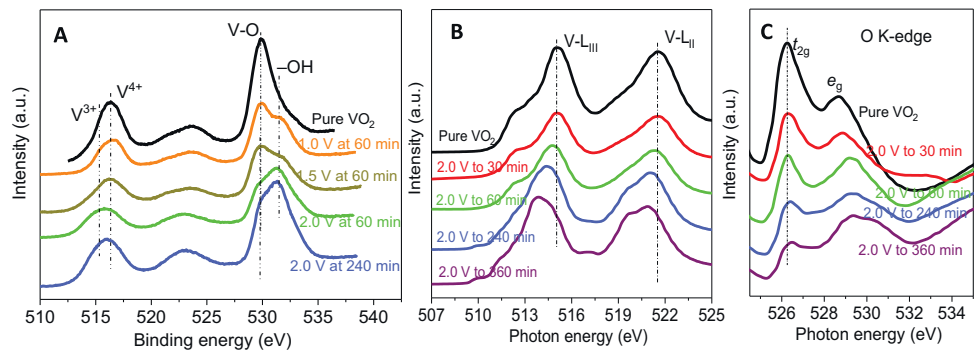

**Fig. 3. Chemical and electronic states of gating-treated VO$_2$ films.** (**A**) XPS curves of VO$_2$ films with different gating voltages. (**B**) V L-edge XANES curves and (**C**) O K-edge curves for the samples gated at 2.0 V with different gating times.

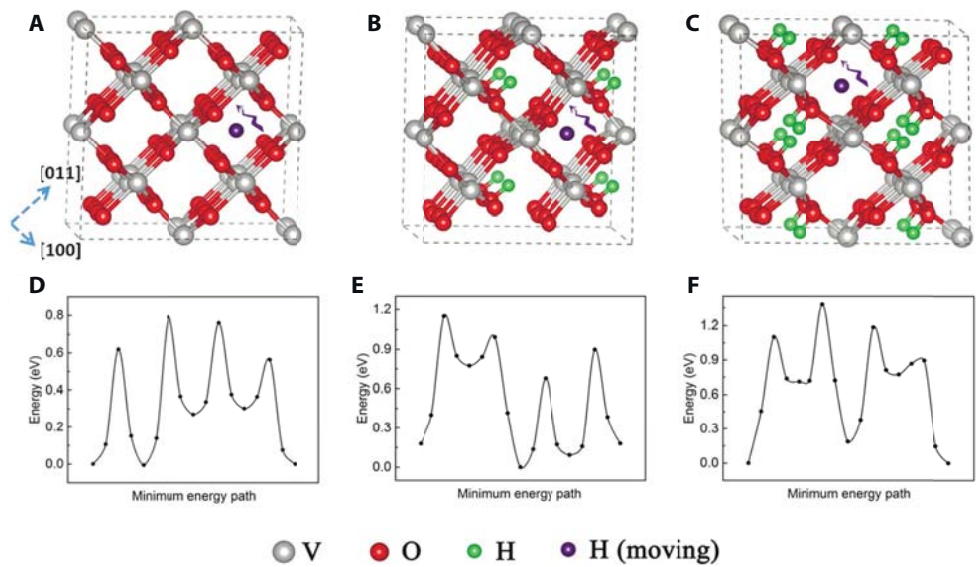

**Fig. 4. Theoretical calculation for H insertion and diffusion processes.** (**A** to **C**) Optimized atomic structures of V$_4$O$_8$ (pure VO$_2$), HV$_4$O$_8$, and H$_2$V$_4$O$_8$, respectively. Purple beads with arrows represent H-diffusion channels along the [100] direction. (**D** to **F**) Calculated energy barriers for diffusion of H atoms into V$_4$O$_8$, HV$_4$O$_8$, and H$_2$V$_4$O$_8$, respectively.

## Gate-controlled VO$_2$-based smart window with high performance

The smart window concept was proposed many years ago, and many materials showed promise for use as smart windows, including hydrogels, liquid crystals, transition metal oxides, and perovskite (41, 42), because of their special optical response in the visible region upon external stimulus/excitation. The VO$_2$ material exhibited a special switching effect in the near-infrared (NIR) range across the phase transition, which demonstrated a unique advantage in energy-saving applications. Since the phase transformations of VO$_2$ films can be controlled by gating treatment at room temperature, we were able to apply it to a voltage-controlled smart window, as shown schematically in Fig. 5A. For the VO$_2$ films with three different phase structures, obvious color changes were observed. The pure VO$_2$ film and the gating-produced metallic H$_x$VO$_2$ film showed a similar yellowish color, while the heavily H-doped VO$_2$ film was almost transparent. In Fig. 5B, the metallic H$_x$VO$_2$ film showed transmittance spectra similar to those of the rutile VO$_2$ film (VO$_2$ film at 90°C), demonstrating a pronounced switching-off effect in the infrared region. Conversely, the insulating HVO$_2$ film maintained high transparency in both visible light and infrared regions. For example, the transmittance at 650 nm for the pristine VO$_2$ film and metallic

H$_x$VO$_2$ was about 56%, while that of the insulating HVO$_2$ film was increased up to 72%. In the infrared region, the transmittance difference at 2000 nm for the pure VO$_2$ film across the MIT process (25° and 90°C) was about 40.3%, while the change from the metallic H$_x$VO$_2$ film to the insulating HVO$_2$ film reached 49.1%.

On the basis of these spectral measurements, the integrated luminous transmittances (380 to 780 nm) and solar transmittances (350 to 2500 nm) can be calculated by the following equations (18)

$$T_{lum} = \int \Phi_{lum}(\lambda) T(\lambda) d\lambda / \int \Phi_{lum}(\lambda) d\lambda \qquad (3)$$

$$T_{sol} = \int \Phi_{sol}(\lambda) T(\lambda) d\lambda / \int \Phi_{sol}(\lambda) d\lambda \qquad (4)$$

where $T(\lambda)$ is the transmittance function, $\Phi_{lum}$ is equal to the standard luminous efficiency function for photopic vision, and $\Phi_{sol}$ represents the solar irradiance spectrum for an air mass of 1.5. The solar modulation ability ($\Delta T_{sol}$) of the metallic H$_x$VO$_2$ and insulating HVO$_2$ films can be calculated by

$$\Delta T_{sol} = T_{sol,insulating-HVO_2} - T_{sol,metallic-H_{1-x}VO_2} \qquad (5)$$

According to the above formula, the transparent, heavily insulating HVO$_2$ film exhibits a high $T_{lum}$ value of about 70.8%, while the transition between the metallic and insulating phases of the H-doped VO$_2$ film is accompanied by a good $\Delta T_{sol}$ value of 26.5% (Fig. 5C).

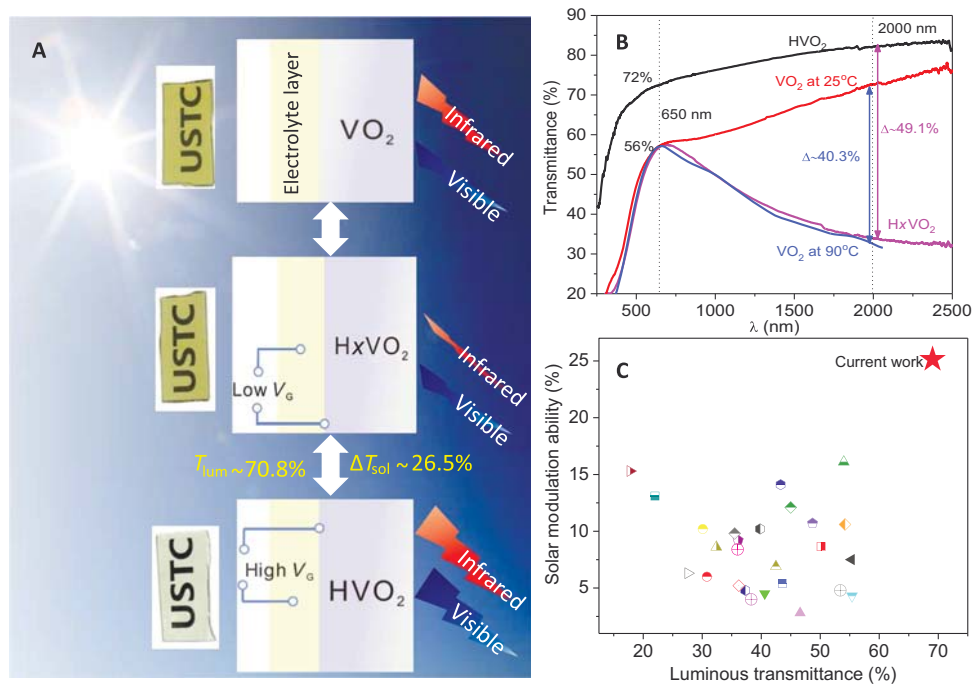

**Fig. 5. Optical properties in relation to possible smart window applications.** (**A**) The schematic diagram of an electrochromic smart window. Properties of three films with different phases are compared. Both the pure VO$_2$ film and the gating-produced metallic H$_x$VO$_2$ film show a similar yellowish color, while the HVO$_2$ film is almost transparent. USTC, University of Science and Technology of China. (**B**) Optical transmittance spectra of VO$_2$, H$_x$VO$_2$, and HVO$_2$ thin films. The transmittance at 650 nm for the pristine VO$_2$ film and metallic H$_x$VO$_2$ is about 56%, while the insulating HVO$_2$ film has an increased transmittance of 72%. In the infrared region, the transmittance difference at 2000 nm for the pure VO$_2$ film across the MIT process (transmittance of film at 25°C compared to 90°C) is about 40.3%, while the change in transmittance from the metallic H$_x$VO$_2$ film to the insulating HVO$_2$ film is higher at 49.1%. (**C**) A plot of luminous transmittance $T_{lum}$ versus solar modulation ability $\Delta T_{sol}$ that highlights the results of this work (★) compared to previously reported data.

The data from previous reports for traditional VO$_2$ smart windows (see table S2) are also plotted. The performance of a smart window based on the transition between the metallic and insulating phases of an H-doped VO$_2$ film appreciably surpasses previous records. Furthermore, it even surpasses the theoretical limitation of the $\Delta T_{sol}$ of ~23% for a traditional VO$_2$ smart window, which can only be achieved by texture surface design or complex multiyear film fabrication (43).

For architectural smart coatings, the excellent luminous transmittance in the visible range and high modulation efficiency in the NIR regions are both important and greatly desirable. There are two long-lasting issues hindering the practical application of VO$_2$ in smart windows: low luminous transmittance ($T_{lum}$) and poor solar energy regulation ability ($\Delta T_{sol}$). For a traditional smart window constructed with the pure VO$_2$ film, increasing the VO$_2$ film thickness is necessary for better infrared switching performace, and as a result, the visible light transmittance will be greatly decreased. This problem is caused by the intrinsic band structures and optical properties of VO$_2$, making it very difficult to be solved for the theromochromic VO$_2$ smart window. In addition, its relatively high critical temperature (~68°C) is another big obstacle for practical applications (14, 16). Thus, in this work, the voltage-controlled phase transitions between the metallic H$_x$VO$_2$ and insulating HVO$_2$ films not only overcome the temperature issue by operating at room temperature but also demonstrate high $T_{lum}$ and large $\Delta T_{sol}$ values in Fig. 5C, showing the unique advantages for practical smart window applications.

## DISCUSSION
In this study, we have achieved a reversible tristate phase transformation in a VO$_2$ epitaxial film by using solid electrolyte–assisted electric field control. These gating voltage–controlled phase transitions originate from the insertion/extraction of hydrogens into/from the VO$_2$ lattice. By changing the voltage, we can reversibly modulate the phase transformations among the insulating monoclinic VO$_2$, the metallic H$_{1-x}$VO$_2$ (0 < x < 1), and the insulating HVO$_2$ phase structures. The transition between the metallic H$_x$VO$_2$ phase and the insulating HVO$_2$ phase shows pronounced electrochromic effects at room temperature with greatly enhanced luminous transmittance in the visible range and the highest modulation efficiency in the NIR regions. This system overcomes the primary weaknesses of the traditional VO$_2$ smart window. Our findings thus demonstrate a facile route using phase modulation by a solid electrolyte for creating realistic energy-saving devices in the future.

## MATERIALS AND METHODS
### Thin-film growth
High-quality VO$_2$ (020) epitaxial films were grown on an Al$_2$O$_3$ (0001) crystal substrate by radio frequency (rf) plasma–assisted oxide molecular beam epitaxy equipment; more details on film preparation were reported elsewhere (44).

### Solid electrolyte preparation
NaClO$_4$ (Sigma-Aldrich) dissolved in a PEO [molecular weight ($M_w$) = 100,000; Sigma-Aldrich] matrix was used as the solid electrolyte. NaClO$_4$ and PEO powders (0.3 and 1 g, respectively) were mixed with 15 ml of anhydrous methanol (Alfa Aesar). The mixture was then stirred overnight at 50°C until a gel-like solid electrolyte was formed (45). A 10 mm by 10 mm electrolyte-covered VO$_2$ thin-film FET structure with three terminals was fabricated.

## Characterizations
Electrical properties were conducted by Keithley 2400 sourcemeters at room temperature. For the gating test, the voltage sweep rates were set to 1.0 V/2500 s for the fast mode and 1.0 V/10,000 s for the slow mode. Raman spectra (JY LabRAM HR, Ar+ laser at 514.5 nm, 0.5 mW) were recorded for different samples at room temperature. The laser spot size was 1 μm by 2 μm, and five points were tested for each film sample, which showed consistent results. The XPS (Thermo ESCALAB 250; Al $K\alpha$, 1486.7 eV) was used to examine the chemical states of the VO$_2$ film samples before and after gating treatment. To detect the hydrogen concentration in the gated VO$_2$ film sample, SIMS measurements (Quad PHI 6600) were conducted.

XANES was conducted at the x-ray magnetic circular dichroism (XMCD) beamline (BL12B) at the National Synchrotron Radiation Laboratory (NSRL), Hefei. The NSRL is a second-generation accelerator with a 0.8-GeV storage ring, which is suitable for soft x-ray experiments. Total electron yield mode was applied to collect the sample drain current under a vacuum higher than $3.0 \times 10^{-10}$ torr. The energy range is 100 to 1000 eV with an energy resolution ($\Delta E/E$) of $1.0 \times 10^{-3}$ at 1.0 keV, and the beam size is about 1.0 mm by 3.0 mm. High-resolution synchrotron XRD curves were characterized at the BL14B beamline of the Shanghai Synchrotron Radiation Facility (SSRF). The SSRF is a third-generation accelerator with a 3.5-GeV storage ring. The BL14B beamline shows an energy resolution ($\Delta E/E$) of $1.5 \times 10^{-4}$ at 10 keV and a beam size of 0.3 mm by 0.35 mm with a photon flux of up to $2 \times 10^{12}$ phs/s at 10 keV. Considering the photon flux distribution and the resolution, the 0.124-nm x-ray was chosen for the experiment.

### First-principles calculations
All calculations were based on DFT, using the Vienna ab initio Simulation Package code (46). Exchange and correlation terms were described using general gradient approximation (GGA) in the scheme of Perdew-Burke-Ernzerhof (47). Core electrons were described by pseudopotentials generated from the projector-augmented wave method (48), and valence electrons were expanded in a plane-wave basis set with an energy cutoff of 480 eV. The GGA + U method was used to optimize the structure; U and J were chosen to be 4 and 0.68 eV, respectively. The relaxation was carried out until all forces on the free ions converged to 0.01 eV/Å. The climbing image nudged elastic band (49) method was used to find the minimum energy paths and the transition states for diffusion of H into VO$_2$ and H-doped VO$_2$.

**Acknowledgments**
**Funding:** This work was partially supported by the National Key Research and Development Program of China (2018YFA0208603, 2016YFA0401004), the National Natural Science Foundation of China (U1432249, 11404095, 11574279, and 11704362), the Fundamental Research Funds for the Central Universities, the Youth Innovation Promotion Association CAS, the Major/Innovative Program of Development Foundation of Hefei Center for Physical Science and Technology, and the China Postdoctoral Science Foundation (2017M62002). This work was partially carried out at the USTC Center for Micro- and Nanoscale Research and Fabrication. We also acknowledge support from the XMCD beamline (BL12B), the photoelectron spectroscopy beamline (BL10B) at the NSRL (Hefei), and the x-ray diffraction beamline (BL14B1) at the SSRF. **Author contributions:** J.J. and C.Z. conceived the study. S.C. and C.Z. designed the experiments and performed the initial tests. Z.W. and J.J. conducted the theoretical calculations. S.C., H.R., Y.C., W.Y., C.W., and B.L. conducted the sample characterizations and synchrotron-based measurements. S.C., J.J., and C.Z. wrote the manuscript. All authors discussed the results and commented on the manuscript. **Competing interests:** The authors declare that they have no competing interests. **Data and materials availability:** All data needed to evaluate the conclusions in the paper are present in the paper and/or the Supplementary Materials. Additional data related to this paper may be requested from the authors.

Submitted 10 October 2018
Accepted 30 January 2019
Published 15 March 2019
10.1126/sciadv.aav6815

**Citation:** S. Chen, Z. Wang, H. Ren, Y. Chen, W. Yan, C. Wang, B. Li, J. Jiang, C. Zou, Gate-controlled $VO_2$ phase transition for high-performance smart windows. *Sci. Adv.* **5**, eaav6815 (2019).

# Science Advances

## Gate-controlled VO$_2$ phase transition for high-performance smart windows

Shi Chen, Zhaowu Wang, Hui Ren, Yuliang Chen, Wensheng Yan, Chengming Wang, Bowen Li, Jun Jiang and Chongwen Zou

| | |
|---|---|
| ARTICLE TOOLS | http://advances.sciencemag.org/content/5/3/eaav6815 |
| SUPPLEMENTARY MATERIALS | http://advances.sciencemag.org/content/suppl/2019/03/11/5.3.eaav6815.DC1 |
| REFERENCES | This article cites 58 articles, 4 of which you can access for free http://advances.sciencemag.org/content/5/3/eaav6815#BIBL |
| PERMISSIONS | http://www.sciencemag.org/help/reprints-and-permissions |

Use of this article is subject to the Terms of Service